\documentclass[twocolumn]{jpsj3}
\usepackage{graphicx,color}

\usepackage{color}
\usepackage{textcomp}
\usepackage{bm}

\def\runauthor{Online-Journal Subcommittee}

\title{%
Structural Analysis on Iron-Based Superconductor Pr1111 System with Oxygen Deficiency and Flourine Substitution}

\author{%
Katsuaki \textsc{Kodama}$^{1,2}$, Motoyuki \textsc{Ishikado}$^{1,2}$, Fumitaka \textsc{Esaka}$^{3}$, 
Akira \textsc{Iyo}$^{2,4}$, Hiroshi \textsc{Eisaki}$^{2,4}$, and Shin-ichi \textsc{Shamoto}$^{1,2}$}

\inst{%
$^{1}$Quantum Beam Science Directorate, Japan Atomic Energy Agency, Tokai, Ibaraki 319-1195, Japan \\
$^{2}$JST, Transformative Research-Project on Iron Pnictides (TRIP), Tokyo 102-0075, Japan \\
$^{3}$Nuclear Science and Engineering Directorate, Japan Atomic Energy Agency, Tokai, Ibaraki 319-1195, Japan \\
$^{4}$Nanoelectronics Research Institute, National Institute of Advanced
   Industrial Science and Technology, Tsukuba, Ibaraki 305-8562, Japan \\
}

\recdate{\today}

\abst{%
We have performed structural analyses on iron-based superconductors, PrFeAsO$_{1-y}$ and PrFeAsO$_{1-x}$F$_x$, 
systematically, by means of Rietveld method on neutron powder diffraction data.  
The shifts of iron ion valence from +2, $\delta$, are accurately determined from the occupancies of O 
and O$_{1-x}$F$_x$ sites obtained by the Rietveld analysis and F-concentration 
obtained by secondary ion-microprobe mass spectrometry .  
$T_\textrm{c}$-$\delta$ curve of PrFeAsO$_{1-y}$ is different from the curve of PrFeAsO$_{1-x}$F$_x$,  
indicating that $\delta$ is not a principal parameter for $T_\textrm{c}$ in so-called 1111 system.  
Structural parameters of the FeAs layers, for example, As-Fe-As bond angle and As-height from Fe layer, 
are different between both systems with similar $\delta$-values.  Their parent compounds are also found to have 
different structural parameters, possibly due to the different synthetic conditions.  
These results suggest that the difference of structural parameters of FeAs layer is the origin of the discrepancy of 
$T_\textrm{c}$-$\delta$ curves of both systems and the $T_\textrm{c}$-value in the 1111 system is sensitive to 
the structural parameters.  
It may be attribute to an energy balance of the conducting bands contributing to the superconductivity.  } 

\kword{iron-based superconductor, structural analysis, neutron diffraction}

\begin{document}
\maketitle

\section{Introduction}
Iron-based high-$T_\textrm{c}$ superconductor was first discovered by the partial substitution of flourine for 
oxygen in LaFeAsO which is a semimetal with an antiferromagnetic ordering.\cite{kamihara}  
After this dicovery, the series of RFeAsO$_{1-x}$F$_x$ systems were synthesized and the superconducting transition
temperature $T_\textrm{c}$ reached up to about 55~K, where R is lanthanide elements.\cite{chen1, chen2, ren, chen3, liu}
In these systems, the partial substitution of flourine for oxygen causes the change of the valence of iron ion from 
+2 to +(2-x), resulting in so-called electron doping in the conducting bands which consist of Fe 3$d$ orbitals.  
The electron doping is considered to cause the phase transition from antiferromagnetic state to superconducting 
state.  In term of same idea, RFeAsO$_{1-y}$ systems with the oxygen deficiency which also induces the electron doping 
are synthesized under high pressure and they also exhibit the superconductivity.\cite{kito}  
Other compounds, AFe$_2$As$_2$ with A=Ba or Sr, show the transition from antiferromagnetic ordering to superconducting state 
by the substitution of Co for Fe, which gives the excess 3$d$ electrons.\cite{sefat} 
\par
In early stage of iron-based superconductor research, theoretical studies point out the role of electron doping.  
The 3$d$ orbitals of Fe give the Fermi surface of a hole at $\Gamma$-point and the Fermi surface of an electron 
at M-point in reduced Brilliouin zone.\cite{singh}     
At non-doping, the nesting between Fermi surfaces at $\Gamma$- and M-points induces the spin density wave state (SDW) 
or antiferromagnetic ordering state.  
The electron doping suppresses the SDW, and, however, the nesting gives a dynamical spin fluctuation which is considered 
to be origin of the superconductivity.\cite{mazin,cvetkovic,kuroki1,ma}
The recent experimental results support the above theoretical picture.  
The inelastic neutron scattering measurements on LaFeAsO$_{1-x}$F$_x$ with $x$=0.057, 0.082 and 0.157 indicate that 
the spin fluctuation observed in the sample with $x$=0.057 and 0.082 which exhibit superconductivity, 
almost disappears in the sample with $x$=0.157 in which the high-$T_\textrm{c}$ superconductivity is 
almost suppressed.\cite{wakimoto}
It can be explained that the excess electron doping shrinks the hole Fermi surface at $\Gamma$-point.  
The shrinking of the hole Fermi surface is also observed by ARPES on over-doped BaFe$_{2-x}$Co$_x$As$_2$ with $x$=0.3 
in which the superconductivity is almost suppressed.\cite{sekiba}  
These results show the importance of the electron doping and/or the valence shift of Fe ion from +2 
to the superconducivity.
\par
However, several experimental results indicate that the valence shift of Fe ion is not solitary parameter to 
control the electronic state in the iron-based superconductors.  
$T_\textrm{c}$ of LaFeAsO$_{1-x}$F$_x$ system is highly suppressed around $x$=0.20, while the $T_\textrm{c}$ of 
other lanthanide systems, for example, Ce and Pr, remain near maximum values at similar $x$-region.\cite{luetkens, zhao, rotundu}  
The Fe-valence at the boundary between antiferromagnetic ordering and superconducting phases in LaFeAsO$_{1-x}$F$_x$ system 
is about +1.95 ($x\sim 0.05$),\cite{kamihara} very different from the value of 
about +1.6 of PrFeAsO$_{1-y}$ system ($y\sim 0.2$).\cite{kito}  
On the other hand, early structural analysis on RFeAsO$_{1-y}$ shows that the maximum $T_\textrm{c}$ 
values of RFeAsO$_{1-y}$ systems depend on As-Fe-As bond angles.\cite{lee}  
Theoretical study indicates that an amplitude and $\bm{Q}$-dependence of spin fluctuation depend on the height of pnictogen ion, 
resulting in the difference of maximum $T_c$-values in RFeAsO$_{1-y}$ systems.\cite{kuroki2}  
These results suggest that the structural parameters of FeAs layer are also important parameters 
for the electronic state in the iron-based superconductors.  
We have performed sturctural analyses on 1111 system with identical lanthanide element, Pr,  
in which superconductivities are induced by oxygen deficiency and flourine substitution,  
in order to obtain information on the role of structural parameters to the electronic state and superconductivity.  
\par

\section{Experiments}
Powder samples of PrFeAsO$_{1-y}$ and PrFeAsO$_{1-x}$F$_{x}$ were prepared by following processes.  
Polycrystalline samples of PrFeAsO$_{1-y}$ have been synthesized by high pressure method using belt-type-anvil apparatus. 
Powders of PrAs, Fe, Fe$_2$O$_3$ were used as the starting materials for PrFeAsO$_{1-y}$. 
PrAs was obtained by reacting Pr powders and As grains at 500 \textcelsius ~for 10 hours and then 850 \textcelsius ~for 5 hours 
in an evacuated quartz tube. 
Mixed starting materials with their nominal compositions of PrFeAsO$_{1-y}$ were sintered at 1000-1200 \textcelsius ~for 2 hours 
under a pressure of about 2~GPa using belt-type-anvil apparatus.  
The samples of about 1.8 g were prepared for neutron diffraction.  
Powder x-ray diffraction data show that obtained samples contain impurities of Pr$_2$O$_3$, FeAs and PrAs.  
The samples with nominal $y$-values larger than 0.2 are not used for the structural analysis because they 
contain the large amount of impurities.  
Powders of PrAs, Fe, Fe$_2$O$_3$, and FeF$_2$ were used as starting materials for the synthesis of PrFeAsO$_{1-x}$F$_{x}$.  
The starting materials with nominal composition of PrFeAsO$_{1-x}$F$_{x}$, 
were mixed and sintered at 1100~\textcelsius~for 10 hours in an evacuated quartz tube.  
The masses of samples for neutron diffraction are about 5 g.  
Powder x-ray diffraction data show that obtained samples contain only single PrFeAsO$_{1-x}$F$_{x}$ phase.
The ratios of oxygen and flourine in the samples with finite $x$ were determined by secondary ion-microprobe mass 
spectrometry (SIMS).  
\par
The $T_\textrm{c}$ values are determined by the eletrical resisitivity measured by PPMS 
and the superconducitng shielding diamagnetism measured by SQUID magnetometer. 
\par
Neutron powder diffraction measurements were performed at room temperature, 
using the high-resolution powder diffractometer HRPD 
(neutron wave length: 1.8234 $\mathrm{\AA}$, collimations: open (effective value of 35')-20'-6') 
installed in the reactor JRR-3 of JAEA.  
Diffraction data were collected with constant monitor counts and a step angle of 0.05~\textdegree~ 
over the 2$\theta$ range of 10-162.4~\textdegree.  
These experimental conditions and the sample positions are completely common to all measurements.  
The powder samples were set in vanadium holders with diameters of 6 and 10 mm for PrFeAsO$_{1-y}$ and PrFeAsO$_{1-x}$F$_{x}$, 
respectively.  

\section{Experimental Results}
Temperature ($T$) dependences of the resistivities of PrFeAsO$_{1-y}$ are shown in Fig. 1(a).  
\begin{figure}[tbh]
\centering
\includegraphics[width=7.5 cm]{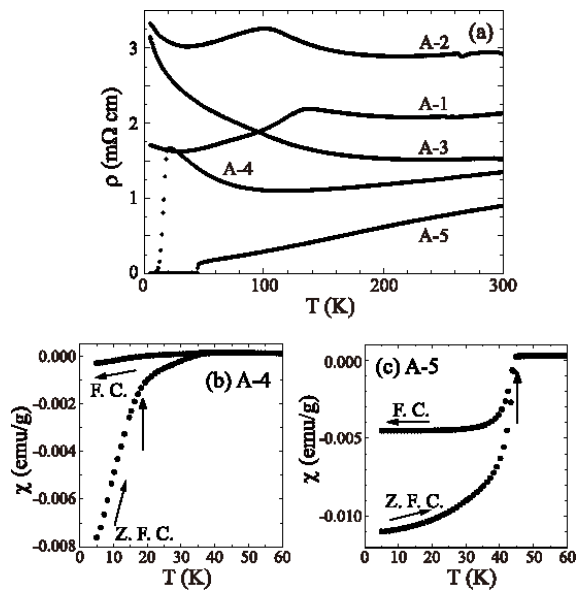} 
\caption{(a) Temperature dependence of the electrical resistivities of PrFeAsO$_{1-y}$ 
with $y$=0.0 (sample A-1), 0.05 (A-2), 0.09 (A-3), 0.10 (A-4) and 0.20 (A-5). For all samples, $y$ are nominal values.  
The magnetic susceptibilities are plotted for superconducting samples, A-4 (b) and A-5 (c).  Signals in zero-field-cooling 
(Z. F. C. : superconducting shielding signal) and field-cooling (F. C. : Meissner signal) are shown.  
Vertical arrows show the onset temperature of superconducting shielding diamagnetism.}
\label{fig.1}
\end{figure}
\begin{figure}[tbh]
\centering
\includegraphics[width=7.5cm]{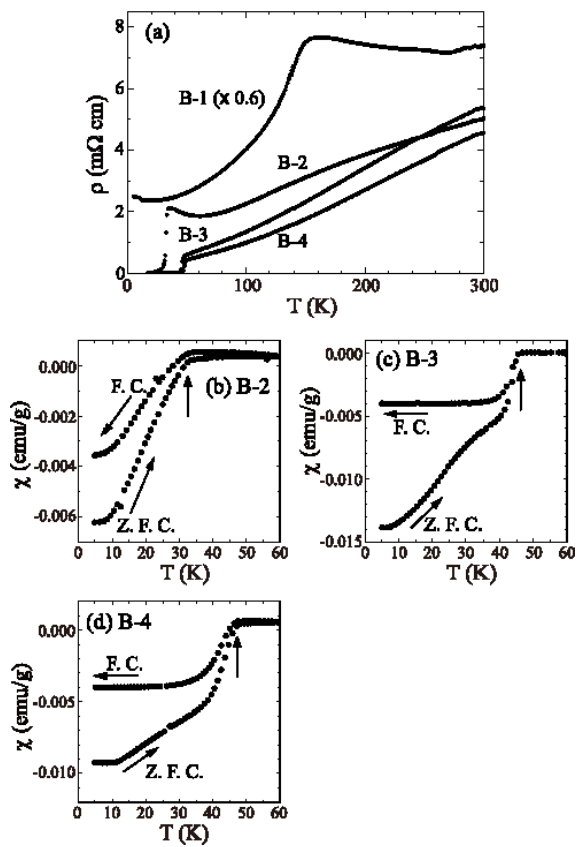} 
\caption{(a) Temperature dependence of the electrical resistivities of PrFeAsO$_{1-x}$F$_x$ 
with x=0.0 (B-1), 0.05 (B-2), 0.10 (B-3) and 0.15 (B-4).  For all samples, $x$ are nominal values.  
The magnetic susceptibilities are plotted for superconducting samples, B-2 (b), B-3 (c) and B-4 (d).  
Signals in zero-field-cooling (Z. F. C. : superconducting shielding signal) and field-cooling (F. C. : Meissner signal) 
are shown.  
Vertical arrows show the onset temperature of superconducting shielding diamagnetism.}
\label{fig.2}
\end{figure}
As shown in Fig. 1(a), the decrease of the resistivity of sample A-4 is somewhat broad at the superconducting transition. 
Here, we define the middle point between the onset of the resistivity drop and 
the zero-resistivity as the $T_\textrm{c}$-value estimated from the resisitivity measurement, $T_\textrm{c}^\textrm{res}$.  
$T$-dependences of the magnetic susceptibilities of superconducting samples A-4 and A-5 are shown in Figs. 1(b) and 1(c).  
The onset temperature of the shielding is defined as $T_\textrm{c}$-value estimated from the shielding measurement, 
$T_\textrm{c}^\textrm{sh}$.  For sample A-4, although the superconducting signal is observed at about 34~K, 
the signal is very small.  Then we define $T_\textrm{c}^\textrm{sh}$=18~K where clear drop of the signal is observed, 
as shown by vertical arrow.  
The values of $T_\textrm{c}^\textrm{sh}$ almost correspond with the values of $T_\textrm{c}^\textrm{res}$.  
The resistiviy of the sample with nominal $y$=0.09 (sample A-3) does not exhibit superconductivity 
although the anomaly caused by the antiferromagnetic ordering and structural phase transition from the tetragonal to 
orthorhombic structure is not observed.  The superconducting shielding diamagnetism is observed below about 16 K.  
However, because the volume fraction of the shielding signal is less than 1 \% at about 4 K, 
this sample is regarded as non-superconducting sample.  
\par
Figure 2(a) shows the $T$-dependences of the resistivities of PrFeAsO$_{1-x}$F$_{x}$.  
In Figs 2(b), 2(c) and 2(d), $T$-dependences of the magnetic susceptibilities of superconducting samples B-2, B-3 and B-4 
are shown, respectively.  
In the case of this system, the superconducting transitions observed in the resistivity and the shielding signal are sharp 
and the values of $T_\textrm{c}^\textrm{res}$ almost correspond with the values of $T_\textrm{c}^\textrm{sh}$.  
\par
Figures 3(a) and 3(b) show the neutron powder diffraction patterns of the samples with nominal compositions of PrFeAsO$_{0.9}$ 
(sample A-4) and PrFeAsO$_{0.9}$F$_{0.1}$ (sample B-3), 
as typical examples of the data on PrFeAsO$_{1-y}$ and PrFeAsO$_{1-x}$F$_{x}$, respectively.  
\begin{figure}[tbh]
\centering
\includegraphics[width=8.5cm]{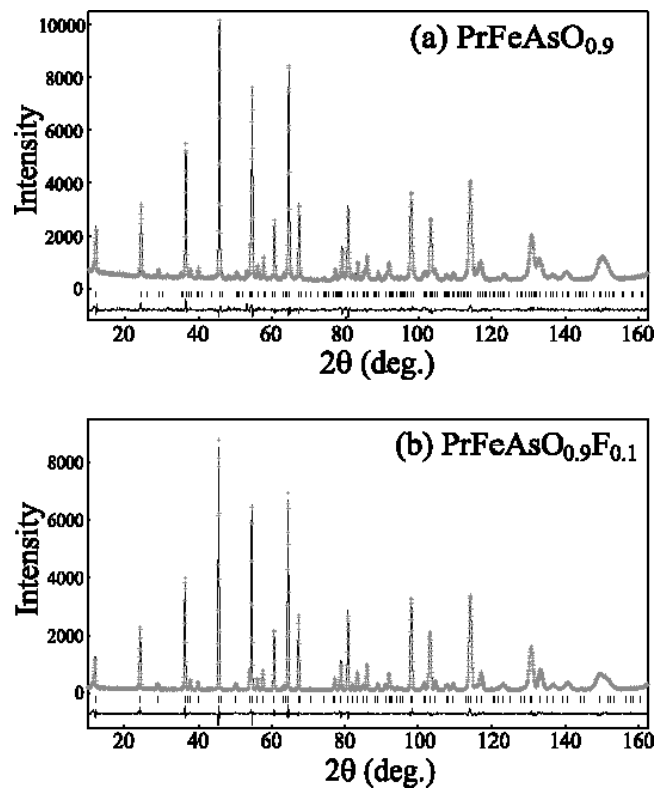} 
\caption{Observed (crosses) and calculated (solid lines) neutron powder diffraction patterns of the samples with nominal 
compositions of PrFeAsO$_{0.9}$ (a) and PrFeAsO$_{0.9}$F$_{0.1}$ (b), respectively.  
Vertical bars show the calculated position of Bragg reflections including 
the impurities.  The solid lines at the bottom of the figures are the differences between observed and calculated intesities}
\label{fig.3a}
\end{figure}
The observed data are shown by crosses.   
Structural analyses on neutron powder diffraction data are performed by using the program RIETAN2000.\cite{izumi}  
The space group of $P4/mmm$ is used.  
The data of PrFeAsO$_{1-y}$ system are analyzed including Pr$_2$O$_3$, FeAs and PrAs as impurities, 
while the data of PrFeAsO$_{1-x}$F$_{x}$ system are analyzed as single phase samples.  
The occupation factor of O and O$_{1-x}$F$_{x}$ sites of PrFeAsO$_{1-y}$ and PrFeAsO$_{1-x}$F$_{x}$ 
are also refined, respecviely, in order to determine the valence of Fe ion accurately.  
The obtained structural parameters of PrFeAsO$_{1-y}$ and PrFeAsO$_{1-x}$F$_{x}$ are shown in Tables I and II, 
respectively.  
Errors of the parameters shown in the tables are mathematical standard deviations obtained by Rietveld analysis. 
Mass fractions of the impurities in PrFeAsO$_{1-y}$ are also shown in Table I.  
The diffraction patterns calculated by using refined parameters are shown in Figs. 3(a) and 3(b) by solid lines.  
The calculated lines can reproduce the observed data.  
In Table II, the accurate O$_{1-x}$F$_{x}$ ratios of PrFeAsO$_{1-x}$F$_{x}$ samples estimated by SIMS are also shown. 
\begin{table}
\caption{Structural parameters of PrFeAsO$_{1-y}$ (space group $P4/mmm$) determined by Rietveld analyses 
of neutron powder diffraction data at room temperature.  Compositions of starting mixtures with $y$=0.0(a), 
$y$=0.05(b), $y$=0.09(c), $y$=0.10(d), and $y$=0.20(e). $B$ is the isotropic displacement parameter.}
\label{t1}
\begin{center}
\begin{footnotesize}
\begin{tabular}{lllllll}
\hline
Atom & Site & Occ. & $x$ & $y$ & $z$ & $B$ \\
\hline
\multicolumn{7}{l}{(a) Sample A-1 (nominal $y$=0.0, non-super.)} \\
\multicolumn{7}{l}{$a$=3.9861(2), $c$=8.6007(4) \AA, $R_\textrm{wp}=5.67~\%$, $R_\textrm{wp}/R_\textrm{e}=1.40$} \\
Pr & 2c & 1 & 1/4 & 1/4 & 0.1387(4) & 0.50(7) \\
Fe & 2b & 1 & 3/4 & 1/4 & 1/2 & 0.39(4) \\
As & 2c & 1 & 1/4 & 1/4 & 0.6566(3) & 0.66(5) \\
O & 2a & 0.992(6) & 3/4 & 1/4 & 0 & 0.28(6) \\
\multicolumn{7}{l}{impurity : Pr$_2$O$_3$ 9.1$~\%$, FeAs 8.6$~\%$, PrAs 0.0$~\%$} \\
\multicolumn{7}{l}{ } \\
\multicolumn{7}{l}{(b) Sample A-2 (nominal $y$=0.05, non-super.)} \\
\multicolumn{7}{l}{$a$=3.9854(2), $c$=8.5963(3) \AA, $R_\textrm{wp}=5.32~\%$, $R_\textrm{wp}/R_\textrm{e}=1.33$} \\
Pr & 2c & 1 & 1/4 & 1/4 & 0.1394(4) & 0.49(5) \\
Fe & 2b & 1 & 3/4 & 1/4 & 1/2 & 0.46(4) \\
As & 2c & 1 & 1/4 & 1/4 & 0.6562(2) & 0.52(4) \\
O & 2a & 0.972(4) & 3/4 & 1/4 & 0 & 0.26(5) \\
\multicolumn{7}{l}{impurity : Pr$_2$O$_3$ 2.9$~\%$, FeAs 2.3$~\%$, PrAs 0.0$~\%$} \\
\multicolumn{7}{l}{ } \\
\multicolumn{7}{l}{(c) Sample A-3 (nominal $y$=0.09, non-super.)} \\
\multicolumn{7}{l}{$a$=3.9861(1), $c$=8.5884(3) \AA, $R_\textrm{wp}=5.53~\%$, $R_\textrm{wp}/R_\textrm{e}=1.15$} \\
Pr & 2c & 1 & 1/4 & 1/4 & 0.1390(3) & 0.77(4) \\
Fe & 2b & 1 & 3/4 & 1/4 & 1/2 & 0.62(3) \\
As & 2c & 1 & 1/4 & 1/4 & 0.6564(2) & 0.55(3) \\
O & 2a & 0.982(4) & 3/4 & 1/4 & 0 & 0.44(4) \\
\multicolumn{7}{l}{impurity : Pr$_2$O$_3$ 2.5$~\%$, FeAs 1.8$~\%$, PrAs 0.0$~\%$} \\
\multicolumn{7}{l}{ } \\
\multicolumn{7}{l}{(d) Sample A-4 (nominal $y$=0.10, $T_\textrm{c}^\textrm{res}$=15.8 K, $T_\textrm{c}^\textrm{sh}$=18 K)} \\
\multicolumn{7}{l}{$a$=3.9798(1), $c$=8.5880(2) \AA, $R_\textrm{wp}=4.92~\%$, $R_\textrm{wp}/R_\textrm{e}=1.27$} \\
Pr & 2c & 1 & 1/4 & 1/4 & 0.1400(3) & 0.83(6) \\
Fe & 2b & 1 & 3/4 & 1/4 & 1/2 & 0.55(4) \\
As & 2c & 1 & 1/4 & 1/4 & 0.6574(2) & 0.47(4) \\
O & 2a & 0.908(4) & 3/4 & 1/4 & 0 & 0.13(5) \\
\multicolumn{7}{l}{impurity : Pr$_2$O$_3$ 0.0$~\%$, FeAs 2.7$~\%$, PrAs 2.0$~\%$} \\
\multicolumn{7}{l}{ } \\
\multicolumn{7}{l}{(e) Sample A-5 (nominal $y$=0.20, $T_\textrm{c}^\textrm{res}$=45.3 K, $T_\textrm{c}^\textrm{sh}$=45 K)} \\
\multicolumn{7}{l}{$a$=3.9709(2), $c$=8.5839(3) \AA, $R_\textrm{wp}=6.52~\%$, $R_\textrm{wp}/R_\textrm{e}=1.54$} \\
Pr & 2c & 1 & 1/4 & 1/4 & 0.1423(3) & 0.60(6) \\
Fe & 2b & 1 & 3/4 & 1/4 & 1/2 & 0.35(4) \\
As & 2c & 1 & 1/4 & 1/4 & 0.6579(3) & 0.52(5) \\
O & 2a & 0.857(6) & 3/4 & 1/4 & 0 & 0.50(6) \\
\multicolumn{7}{l}{impurity : Pr$_2$O$_3$ 0.0$~\%$, FeAs 2.8$~\%$, PrAs 2.3$~\%$} \\
\hline
\end{tabular}
\end{footnotesize}
\end{center}
\end{table}
\begin{table}
\caption{Structural parameters of PrFeAsO$_{1-x}$F$_{x}$ (space group $P4/mmm$) determined by Rietveld analyses 
of neutron powder diffraction data at room temperature.  Compositions of starting mixtures with $x$=0.0(a), 
$x$=0.05(b), $x$=0.10(c), and $x$=0.15(d). O$_{1-x}$F$_{x}$ ratio in the column are estimated by SIMS.}
\label{t1}
\begin{center}
\begin{tabular}{lllllll}
\hline
Atom & Site & Occ. & $x$ & $y$ & $z$ & $B$ \\
\hline
\multicolumn{7}{l}{(a) Sample B-1 (nominal $x$=0.0, non-super.)} \\
\multicolumn{7}{l}{$a$=3.9810(1), $c$=8.6230(2) \AA, $R_\textrm{wp}=6.44~\%$, $R_\textrm{wp}/R_\textrm{e}=1.19$} \\
Pr & 2c & 1 & 1/4 & 1/4 & 0.1391(2) & 0.78(4) \\
Fe & 2b & 1 & 3/4 & 1/4 & 1/2 & 0.58(3) \\
As & 2c & 1 & 1/4 & 1/4 & 0.6569(2) & 0.51(3) \\
O & 2a & 0.987(4) & 3/4 & 1/4 & 0 & 0.52(4) \\
\multicolumn{7}{l}{ } \\
\multicolumn{7}{l}{(b) Sample B-2 (nominal $x$=0.05, $T_\textrm{c}^\textrm{res}$=32.2 K, $T_\textrm{c}^\textrm{sh}$=32 K)} \\
\multicolumn{7}{l}{$a$=3.9773(1), $c$=8.6108(2) \AA, $R_\textrm{wp}=6.11~\%$, $R_\textrm{wp}/R_\textrm{e}=1.14$} \\
Pr & 2c & 1 & 1/4 & 1/4 & 0.1409(2) & 0.76(4) \\
Fe & 2b & 1 & 3/4 & 1/4 & 1/2 & 0.56(3) \\
As & 2c & 1 & 1/4 & 1/4 & 0.6571(2) & 0.51(3) \\
O$_{0.945}$F$_{0.055}$ & 2a & 0.986(4) & 3/4 & 1/4 & 0 & 0.58(4) \\
\multicolumn{7}{l}{ } \\
\multicolumn{7}{l}{(c) Sample B-3 (nominal $x$=0.10, $T_\textrm{c}^\textrm{res}$=46.1 K, $T_\textrm{c}^\textrm{sh}$=46 K)} \\
\multicolumn{7}{l}{$a$=3.9749(1), $c$=8.6007(2) \AA, $R_\textrm{wp}=6.15~\%$, $R_\textrm{wp}/R_\textrm{e}=1.13$} \\
Pr & 2c & 1 & 1/4 & 1/4 & 0.1430(2) & 0.85(4) \\
Fe & 2b & 1 & 3/4 & 1/4 & 1/2 & 0.59(3) \\
As & 2c & 1 & 1/4 & 1/4 & 0.6577(2) & 0.54(3) \\
O$_{0.878}$F$_{0.122}$ & 2a & 0.990(4) & 3/4 & 1/4 & 0 & 0.56(4) \\
\multicolumn{7}{l}{ } \\
\multicolumn{7}{l}{(d) Sample B-4 (nominal $x$=0.15, $T_\textrm{c}^\textrm{res}$=48.7 K, $T_\textrm{c}^\textrm{sh}$=47 K)} \\
\multicolumn{7}{l}{$a$=3.9711(1), $c$=8.5996(2) \AA, $R_\textrm{wp}=6.34~\%$, $R_\textrm{wp}/R_\textrm{e}=1.19$} \\
Pr & 2c & 1 & 1/4 & 1/4 & 0.1443(3) & 0.96(4) \\
Fe & 2b & 1 & 3/4 & 1/4 & 1/2 & 0.59(3) \\
As & 2c & 1 & 1/4 & 1/4 & 0.6583(2) & 0.47(3) \\
O$_{0.868}$F$_{0.132}$ & 2a & 0.985(4) & 3/4 & 1/4 & 0 & 0.54(4) \\
\hline
\end{tabular}
\end{center}
\end{table}

\section{Discussions}
Now we can accurately estimate the shift of the Fe valence from +2, $\delta$,  
because the occupancies of O (O$_{1-x}$F$_x$) site in PrFeAsO$_{1-y}$ (PrFeAsO$_{1-x}$F$_{x}$) and the ratio of O$_{1-x}$F$_x$ 
in PrFeAsO$_{1-x}$F$_{x}$ are accurately determined by neutron diffraction and SIMS measurements, respectively.  
We use the values of $T_\textrm{c}^\textrm{sh}$ as $T_\textrm{c}$-values because $T_\textrm{c}^\textrm{sh}$ almost correspond 
with $T_\textrm{c}^\textrm{res}$, as shown in previous section.  
In Fig. 4, $T_\textrm{c}$ ({$T_\textrm{c}^\textrm{sh}$) of PrFeAsO$_{1-y}$ and PrFeAsO$_{1-x}$F$_{x}$ are plotted 
as functions of $\delta$, by closed circles and squares, respectively.   
\begin{figure}[tbh]
\centering
\includegraphics[width=7cm]{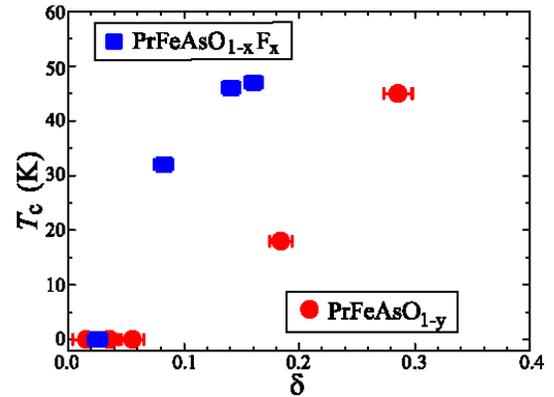} 
\caption{(Color online)$T_\textrm{c}$ of PrFeAsO$_{1-y}$ and PrFeAsO$_{1-x}$F$_{x}$ are plotted as a function of 
$\delta$, by closed circles and squares, respectively.  }
\label{fig.4}
\end{figure}
$T_\textrm{c}$-$\delta$ curve of PrFeAsO$_{1-y}$ obviously deviates from the curve of PrFeAsO$_{1-x}$F$_{x}$.  
$T_\textrm{c}$-value of PrFeAsO$_{1-x}$F$_{x}$ increases against $\delta$ rapidlier than the value of PrFeAsO$_{1-y}$,
and reaches the maximum $T_\textrm{c}$-value at $\delta \sim$ 0.16 
while the $T_\textrm{c}$-value of PrFeAsO$_{1-y}$ almost reaches the maximum at $\delta \sim$ 0.29. 
These results indicate that the $T_\textrm{c}$-value is not determined only by $\delta$-value, at least, in so-called
1111 system even if the system consists of identical lanthanide element. 
\par
The lattice parameters $a$ and $c$ of PrFeAsO$_{1-y}$ and PrFeAsO$_{1-x}$F$_{x}$ are shown in Fig. 5 by the cirlces and squares, 
respectively.  The superconducting and non-superconducting samples are shown by the closed and open symbols.  
The data are plotted as functions of $\delta$.  
\begin{figure}[tbh]
\centering
\includegraphics[width=7cm]{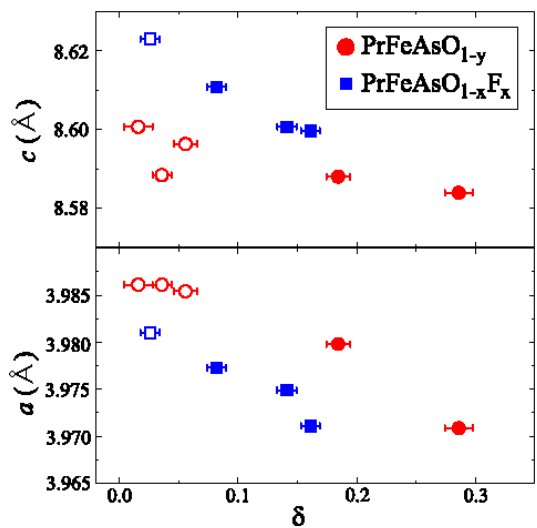} 
\caption{(Color online)Lattice parameters of $a$ (bottom panel) and $c$ (top panel) of PrFeAsO$_{1-y}$ and PrFeAsO$_{1-x}$F$_{x}$
are plotted as a function of $\delta$ by the cirlces and squares, respectively.  
The superconducting and non-superconducting samples are shown by the closed and open symbols.}
\label{fig.5}
\end{figure}
At similar $\delta$-values, the values of $a$ ($c$) of PrFeAsO$_{1-y}$ are larger (smaller) than the values of 
PrFeAsO$_{1-x}$F$_{x}$.  We note that even the parent compounds in both systems corresponding to 
the samples A-1 and B-1 have different lattice parameters although they have very similar composition of 
PrFeAsO$_{0.992}$ and PrFeAsO$_{0.987}$.  This discrepancy of the lattice parameters are caused by the different conditions 
of their syntheses as mentioned in \S 2.  
Because PrFeAsO$_{1-y}$ system including the parent compound are synthesized under high pressure, 
the $c$-axis that PrO and FeAs layers are alternately stacked contracts relative to 
the $c$-axis of PrFeAsO$_{1-x}$F$_{x}$ system.
\par
The difference of lattice parameters naturally causes the difference of the strctural parameters on FeAs layer which
exhibits the superconductivity.  The As-Fe-As bond angle, $\alpha$, defined in ref. 19 is the good indicater of the maximum 
value of $T_\textrm{c}$ in RFeAsO$_{1-y}$ systems.\cite{lee}  
It is suggested that the $T_\textrm{c}$-value becomes maximum as the $\alpha$-value approaches 109.47 \textdegree at 
which the FeAs$_4$ unit is a regular tetrahedron.  
On the other hand, theoretical study suggests that the pnictogen height defined as $(z_\textrm{As}-0.5)\times c$ changes 
the spin fluctuation arsing from Fermi surface nesting between $\Gamma$ and M points which induces the $s\pm$ superconductivity.  
The spin fluctuation arising from the $\Gamma$-M Fermi surface nesting is enhanced with increasing the pnictogen height, 
resulting in the increase of $T_\textrm{c}$.\cite{kuroki2}   
Experimentally, it is suggested that $T_\textrm{c}$ becomes maximum 
as the value of the pnictogen height approaches about 1.38 \AA.\cite{mizoguchi}  
Here we focus on these two structural parameters.  
In Fig. 6(a), $\alpha$-values of PrFeAsO$_{1-y}$ and PrFeAsO$_{1-x}$F$_{x}$ are plotted as functions of $\delta$.     
\begin{figure}[tbh]
\centering
\includegraphics[width=7cm]{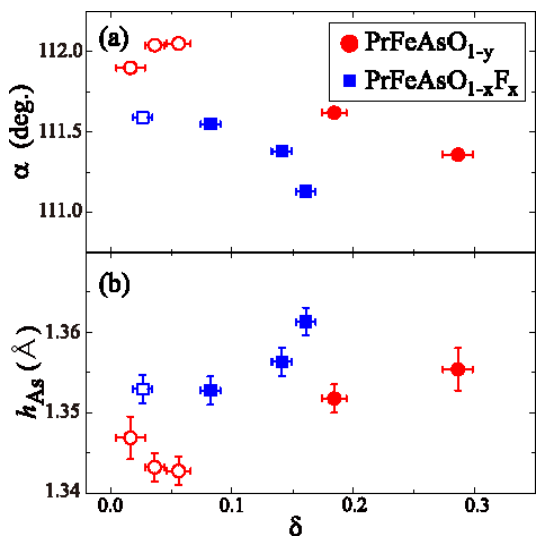} 
\caption{(Color online)As-Fe-As bond angle, $\alpha$, defined in ref. 19 (a) and pnictogen height $h_\textrm{As}$ defined 
in ref. 20 (b) of PrFeAsO$_{1-y}$ and PrFeAsO$_{1-x}$F$_{x}$ are plotted as functions of $\delta$ by the cirlces and squares, 
respectively.  The superconducitng and non-superconducting samples are shown by the closed and open symbols.}
\label{fig.6}
\end{figure}
In both systems, the $\alpha$-values decrease with increasing $\delta$, at least, consistent with the behavior in 
RFeAsO$_{1-y}$ systems reported in ref. 19. 
The $\alpha$-values of PrFeAsO$_{1-y}$ are larger than $\alpha$ of PrFeAsO$_{1-x}$F$_{x}$ at similar $\delta$-values, 
indicating that the FeAs layer of PrFeAsO$_{1-y}$ is flatter than the layer of PrFeAsO$_{1-x}$F$_{x}$ as expected from 
the relation of the lattice parameters between both systems shown in Fig. 5.  In PrFeAsO$_{1-x}$F$_{x}$, 
$\alpha$ reaches the minimum value which almost corresponds with the reported value,\cite{lee, ren2, ren} 
at smaller $\delta$ relative to the case of PrFeAsO$_{1-y}$. In Fig. 6(b), the pnictogen heights, 
$h_\textrm{As}$ of PrFeAsO$_{1-y}$ and PrFeAsO$_{1-x}$F$_{x}$ are plotted as functions of $\delta$.  
The $h_\textrm{As}$-values of PrFeAsO$_{1-x}$F$_{x}$ are larger than $h_\textrm{As}$ of PrFeAsO$_{1-y}$ 
at similar $\delta$-values.  
These results indicate that the structural parameters of FeAs layer are obviously different between 
PrFeAsO$_{1-y}$ and PrFeAsO$_{1-x}$F$_{x}$.  Such difference of FeAs layer may give the difference of 
$T_\textrm{c}$-$\delta$ curves shown in Fig. 4.
\par
The strong correlation between the crystal structure and the magnetism of Fe ion is pointed out by other experimental studies.  
For example, phonon energy contributed from FeAs layer is smaller than the energy obtained by the first principle calculation, 
and this discrepancy can be corrected by the calculation which takes the magnetic moment of Fe into account.\cite{fukuda}  
The amplitude of ordred magnetic moment in the parent compound depend on the pnictgen height.\cite{yin} 
As mentioned in \S1, the correlation between the magnetism and the superconductivity is also pointed out.  
The theoretical studies on iron-based superconductors suggest that three 3$d$ orbitals, 3$d_{x^2-y^2}$, 3$d_{zx}$ 
and 3$d_{yz}$ orbitals form the Fermi surfaces and the spin fluctuation arsing from the nesting 
between the different Fermi surfaces induces the superconductivity.\cite{kuroki1, kurokiunit}  
Based on such senario, the superconductivity and/or the spin fluctuation arising from the Fermi surface nesting can 
be sensitive to the structural parameters of FeAs layer because the energy levels of three 3$d$ orbitals are changed 
by the structural parameters.   
Our results confirm the strong correlation among the structural parameters, the magnetism and the superconductivity 
in the iron-based superconductors.
\par
Here, we note that the change of the lattice parameter $c$ of PrFeAsO$_{1-y}$ 
are not perfectly systematic for the change of $\delta$, as shown in Fig. 5.  
The lattice parameter $c$ of PrFeAsO$_{1-y}$ with $\delta \sim$ 0.04 
corresponding to sample A-3 is obviously smaller than the values of other two non-superconducting samples.  
As a possible origin of such behavior, 
it is suggested that the crystal structure of PrFeAsO$_{1-y}$ system is metastable, at least, at high temperature, 
because PrFeAsO$_{1-y}$ system can be synthesized only under high pressure.  
If the system is metastable, the slight difference of the synthetic condition can cause the difference of structural parameters.  
In addition, the resistivity of the sample A-3 does not exhibit the anomaly caused by antiferromagnetic ordering 
and structural phase transition while the sample A-2 with larger $\delta$-value 
exhibits antiferromagnetic ordering and structural phase transition.  
Such non-systematic behavior to $\delta$ in PrFeAsO$_{1-y}$ system also indicates the strong correlation between the 
structural parameters and the electronic state in iron-based superconductors.  
\par
Finally, we plot the $T_\textrm{c}$-values of PrFeAsO$_{1-y}$ and PrFeAsO$_{1-x}$F$_{x}$ against above structural parameters, 
$\alpha$ and $h_\textrm{As}$.  
In Figs. 7(a) and 7(b), $\alpha$ and $h_\textrm{As}$ dependences of $T_\textrm{c}$ are shown, respectively.  
\begin{figure}[bth]
\centering
\includegraphics[width=7cm]{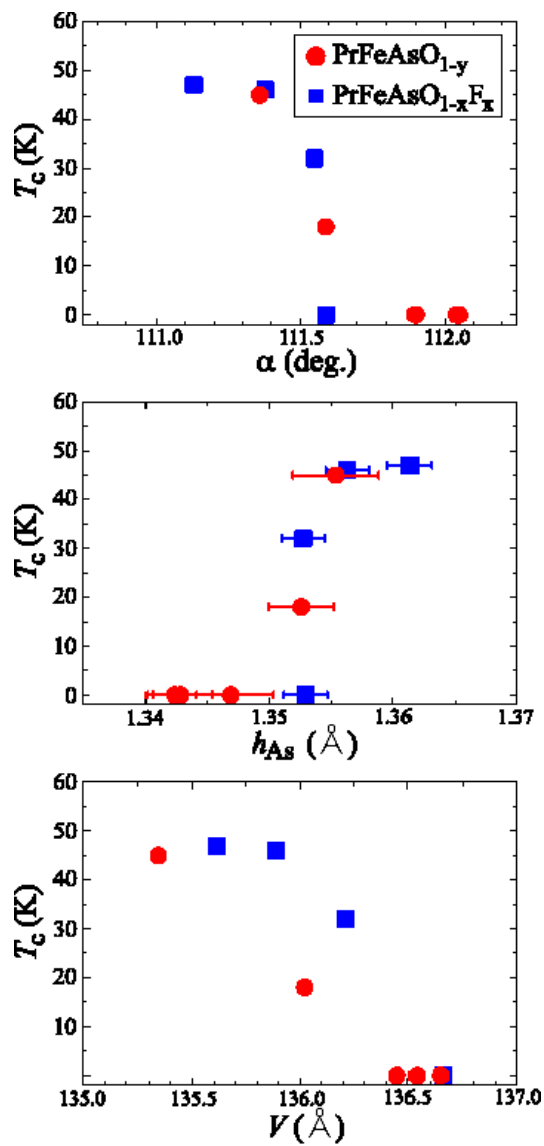} 
\caption{(Color online)$T_\textrm{c}$ of PrFeAsO$_{1-y}$ and PrFeAsO$_{1-x}$F$_{x}$ are plotted by closed circles 
and squares, respectively, as functions of $\alpha$ (a),$h_\textrm{As}$ (b), and volume of unit cell $V$ (c).}
\label{fig.7}
\end{figure}
Note that, in ref. 19 and 22, $T_\textrm{c}$-values of many types of iron-based superconductors whose "carrier doping levels" 
are adjusted to maximum $T_\textrm{c}$, are plotted against the $\alpha$ and $h_\textrm{As}$.  
On the other hand, we plot $T_\textrm{c}$ against the $\alpha$ and $h_\textrm{As}$ from  
antiferromagnetic phase to superconducting phase only on Pr1111 systems with oxygen deficiency and flourine substitution.  
$T_\textrm{c}$-values of PrFeAsO$_{1-y}$ and PrFeAsO$_{1-x}$F$_{x}$ increase 
as $\alpha$-values approach the optimum value of 109.47 \textdegree, 
qualitatively consistent with 1111 systems with other rare earth elements.\cite{lee}  
Similar $\alpha$-dependence of $T_\textrm{c}$ is also observed in hole doped 122 system\cite{rotter}.   
$T_\textrm{c}$-values of PrFeAsO$_{1-y}$ and PrFeAsO$_{1-x}$F$_{x}$ increase 
as $h_\textrm{As}$-values approach the optimum value of 1.38 \AA. 
In FeSe$_{1-x}$Te$_x$ system, the similar behaviors of $\alpha$ and $h_\textrm{As}$ are also observed, at least, 
in the $x$ region from the antiferromagnetic ordering phase at $x$=1.0 to the superconducting phase 
with maximum $T_\textrm{c}$-value at $x$=0.5.\cite{horigane, hosono} 
These results suggest that the $\alpha$ and $h_\textrm{As}$ are important parameters 
for the appearance of the superconductivity although the relation between the $\alpha$ and electronic state is not obvious.
In addition, we also regard the volume-change as important.  
The $T_\textrm{c}$-value increases with decreasing unit cell volume as shown in Fig. 7(c).  
The similar behaviors are also observed in hole- and electron-doped 122 systems,\cite{rotter, canfield} 
and FeSe$_{1-x}$Te$_x$ system in the region of $0.5 \leqq x \leqq 1.0$.
\cite{horigane}  Simply considering the magneto-volume effect, 
a disappearance of the magnetic ordering in the sample with smaller volume is reasonable. 
However, the correlation between $T_\textrm{c}$ and the unit cell volume is not obvious.  
Now we know that the $T_\textrm{c}$-value is sensitive to the structural parameters.  
The further systematic studies to reveal the correlation among the structural parameters, the electronic state 
and the superconductivity are necessary.

\section{Summary}
We have performed the Rietveld analyses on the neutron powder diffraction data of PrFeAsO$_{1-y}$ and PrFeAsO$_{1-x}$F$_{x}$ 
in which the superconductivities are induced by the oxygen deficiency and flourine substitution, respectively.  
The $T_\textrm{c}$-values of both systems are not scaled by the valence shift of Fe ion, $\delta$, 
which is accurately determined by the Rietveld analysis.  
The structural parameters of FeAs layer are different between both systems including the parent compounds 
with the same composition.  
These results suggest that $T_\textrm{c}$-values of the iron-based supercomductors are sensitive to the structural parameters 
of FeAs layer.  
Such sensitivity of $T_\textrm{c}$ to the structural parameters may be due to the energy blance of 3$d$ orbitals contributing 
to the superconductivity which is affected by the structural parameters.  

\section*{Acknowledgment}
The authors would like to thank M. Machida and H. Nakamura for their fruitful discussions and also N. Igawa 
for his help with the neutron diffraction measurements.  
This work is supported by a Grant-in-Aid for Specially Promoted Research 17001001 from 
the Ministry of Education, Culture, Sports, Science and Technology, Japan, and JST TRIP.


\begin{thebibliography}{9}
\bibitem{kurokiunit}
We note that $x$, $y$, and $z$ axes in our text correspond to $X$, $Y$, and $Z$ axes in refs. 12 and 20.
\end{thebibliography}
\end{document}